\documentstyle[prl,aps,floats,epsf]{revtex}

   \renewcommand{\refname}{}
   \newcommand{\biblabel}[1]{[#1]} 
   \renewcommand{\references}{
   \ifpreprintsty

   \vspace*{-0.1 truein}
   \hbox to\hsize{\hss\large \refname\hss}
   \else
   \vskip3pt
   \hrule width\hsize\relax
   \vskip -0.2in
   \fi
   \list{\biblabel{\arabic{enumiv}}}
   {\labelwidth\WidestRefLabelThusFar  \labelsep4pt 
   \leftmargin\labelwidth 
   \advance\leftmargin\labelsep 
   \ifdim\baselinestretch pt>1 pt 
   \parsep  4pt\relax 
   \else 
   \parsep  0pt\relax 
   \fi
   \itemsep\parsep 
   \usecounter{enumiv}
   \def\theenumiv{\arabic{enumiv}}
   }
   \let\newblock\relax 
   \sloppy\clubpenalty4000\widowpenalty4000
   \sfcode`\.=1000\relax
   \ifpreprintsty\else\small\fi
   }

\newcommand{\br}{{\bf r}}

\newcommand{\Vext}{V_{\rm ext}}
\newcommand{\Veff}{V_{\rm eff}}
\newcommand{\Vsc}{V_{\rm sc}}
\newcommand{\FKS}{{\cal F}_{\rm HKS}}
\newcommand{\ETF}{E_{\rm GTF}}
\newcommand{\FTF} {{\cal F}_{\rm GTF}}
\newcommand{\TTF}{{\cal T}_{\rm TF}}
\newcommand{\tTF}{t_{\rm TF}}
\newcommand{\FLDA}{{\cal F}_{\rm DFT}}
\newcommand{\TLDA}{{\cal T}_{\rm DFT}}
\newcommand{\ELDA}{E_{\rm DFT}}
\newcommand{\FH}{F_{\rm H}}
\newcommand{\Eext}{{\cal E}_{\rm ext}}
\newcommand{\Ecoul}{{\cal E}_{\rm coul}}
\newcommand{\Eint}{{\cal E}_{\rm int}}

\newcommand{\Exc}{{\cal E}_{\rm xc}}
\newcommand{\Eop}{{\cal E}_{\rm 1p}}

\newcommand{\Etot}{{\cal E}_{\rm tot}}
\newcommand{\nLDA}{n_{\rm DFT}}
\newcommand{\nLDAW}{n_{\rm DFT}^W}
\newcommand{\nLDAosc}{n_{\rm DFT}^{\rm osc}}
\newcommand{\nTF}{n_{\rm GTF}}
\newcommand{\nosc}{{\tilde n^{\rm osc}}}

\newcommand{\e}{\epsilon}

\newcommand{\muLDAW}{\mu_{\rm DFT}^W}

\begin{document}


\title{Semiclassical Density Functional Theory: \\
Strutinsky Energy Corrections in Quantum Dots} 

\author{Denis Ullmo,$^1$ Tatsuro Nagano,$^2$
Steven Tomsovic,$^2$ and Harold U. Baranger$^3$}

\address{
$^1$Laboratoire~de~Physique~Th\'eorique~et~Mod\`eles~Statistiques~(LPTMS),~91405~Orsay~Cedex,~France\\
$^2$Department of Physics, Washington State University, 
Pullman, WA 99164-2814, USA\\
$^3$Department of Physics, Duke University, Box 90305, 
Durham, NC 27708-0305, USA\\
$\quad$
}

\date{18 July 2000}

\maketitle

\begin{abstract}

We develop a semiclassical density functional theory in the context of
quantum dots.  Coulomb blockade conductance oscillations have been measured
in several experiments using nanostructured quantum dots. The statistical
properties of these oscillations remain puzzling, however, particularly the
statistics of spacings between conductance peaks.  To explore the role that
residual interactions may play in the spacing statistics, we consider
many-body systems which include electron-electron interactions through an
explicit density functional.  First, we develop an approximate series
expansion for obtaining the ground state using the idea of the Strutinsky
shell correction method. Next, we relate the second-order semiclassical
corrections to the screened Coulomb potential.  Finally, we investigate the
validity of the approximation method by numerical calculation of a
one-dimensional model system, and show the relative magnitudes of the
successive terms as a function of particle number.
\\[0.5 cm]

\end{abstract}

\section{Introduction} 

A recurring problem in modern physics is how to add quantization effects to a
basically successful macroscopic theory. This question arises particularly in
the semiclassical regime---large quantum number---where the quantum effects
are often corrections to the essentially classical macroscopic physics.
Perhaps the best known example starts with the Thomas-Fermi theory of the
atom~\cite{qmtext}, which is macroscopic in essence, and then evaluates the
contribution of electronic shell structure to the ground state
energy~\cite{qmtext,march,schwinger}.  The very natural result that the shell
contribution is given by the quantized levels of the self-consistent
Thomas-Fermi potential has been used
extensively~\cite{mayer,latter,bohigas76}.  However, it has only been in
recent decades, starting with the work of V. M.
Strutinsky~\cite{str_str,str_rmp,brack}, that a systematic way of answering
the recurring general problem has been developed.

Our own immediate interest is in quantum dots---small 
electrically conducting regions in which the quantum
properties of the confined electrons are important~\cite{qdot_revs}---and
our aim here is to treat quantum corrections to the ground state energy
of these dots by further developing the Strutinsky method.

Quantum  dots can  be formed,  for instance,  by gate  depletion  of a
two-dimensional electron gas  (2DEG) in a GaAs-AlGaAs heterostructure.
Because of the  high quality of this material  and interface, the mean
free  path  of the  electrons  far exceeds  the  size  of the  quantum
dot. One  can view an  electron as propagating ballistically  within a
confining potential  created by  electrostatic gates patterned  on the
surface of the heterostructure.   For transport measurements, dots can
be coupled  weakly to leads; when  the conductance of  each lead falls
below  $2e^2/h$, electron  transport through  the dot  occurs  only by
tunneling,  and  the  number  of  particles  within  the  dot  becomes
quantized.  In this  regime, the conductance is suppressed  due to the
electrostatic      energy     associated     with      a     localized
charge~\cite{qdot_revs},  an  effect  known  as the  Coulomb  blockade
(first  reported in  1951 by  Gorter \cite{cb_gorter}).  When  the dot
potential is tuned by a gate voltage so that adding one electron costs
no energy, a  large conductance peak appears \cite{qdot_revs}---though
the electrons must still tunnel from the leads, there is no additional
electrostatic  barrier to  conduction.  Sweeping  such a  gate voltage
produces  periodic Coulomb  blockade oscillations  of  the conductance
through dots \cite{qdot_revs,beenakker,vanH}.

The  Coulomb blockade  is a  classical  effect observable  in a  broad
temperature  range,   $k_BT<e^2/C$,  where   $C$  is  the   total  dot
capacitance.  Over most of this  range, both the spacing and height of
the peaks is constant---the spacing is $e^2/C$ and the height is given
by   the  resistance  of   the  two   tunneling  barriers   acting  in
series. However, there is a low temperature regime below a few hundred
millikelvin  for  which  $k_BT<\Delta$,  where $\Delta$  is  the  mean
single-particle  level spacing  of  the isolated  dot. There,  quantum
interference  and  coherence  become important  \cite{qdot_revs}.  The
Coulomb blockade  peaks grow as  the temperature decreases,  and novel
fluctuation  properties   emerge  involving  both   the  peak  heights
\cite{ci_chang,folk,patel}                                          and
spacings~\cite{peakspc_sivan,peakspc_patel,peakspc_simmel1,peakspc_simmel2,peakspc_ensslin}.
The spacings  give information about  the ground state  energies while
the heights involve the magnitude of the wavefunction near the levels.

With some success, random-matrix-theory (RMT) based approaches
\cite{rmt_beenakker} have been used in order first to predict
\cite{ci_jalabert} and then to explain the statistical properties found. In
the simplest approximation \cite{peakspc_sivan}, known as the constant
interaction model, the ground state energy of the dot is expressed as
\[ E(N)=\frac{e^2 N^2}{2C} +\sum_{i=1}^N\epsilon_i \] 
where $N$ is the number of electrons in the dot and $\epsilon_i$ are
single-particle energies. The first term is the classical charging energy;
the second is the total energy of a system of non-interacting
quasi-particles. Supposing that the single-particle classical dynamics within
the dot is chaotic, the Bohigas-Giannoni-Schmit conjecture applies~\cite{bgs}
and implies that the single-particle quantum properties follow RMT. It is
well-established that RMT predicts repulsion amongst the $\epsilon_i$ and
Gaussian random behavior in the eigenfunctions~\cite{brody}. Within this
model, the conductance peak height statistics are in good agreement with
experimental results\cite{ci_chang,folk} after also incorporating non-zero
temperature effects and interference modulations due to periodic paths
coupled to the leads~\cite{ci_jalabert,gok,peakh_narimanov}.
On the other hand, it is 
found~\cite{peakspc_sivan,peakspc_patel,peakspc_simmel1,peakspc_simmel2,peakspc_ensslin}
that the fluctuations in the peak spacings are considerably larger than the
predictions~\cite{peakspc_sivan,blanter,brazil}, and there is no evidence for
the level repulsion or electron spin degeneracy expected from a
single-particle-like approach~\cite{spin_stewart}.  These discrepancies
between the predictions of the constant interaction model and the
observations point to the need for a quantum treatment of the
electron-electron interactions, and, in particular, have triggered a number
of studies based on Hartree-Fock calculations~\cite{walk99,cohen99,ahn99}, or
density functional theory in the local density approximation
(LDA)~\cite{stopa,wingreen}.

In nuclear physics, it has been recognized for some time that the dependence
of nuclear ground state properties on the particle number can be viewed as
the sum of two contributions, a ``smooth part'' which varies smoothly with
the particle number and an ``oscillatory behavior'' producing the shell
structure referred to as shell corrections. A similar decomposition is
possible for any finite-size interacting fermion system. The smooth part
comes basically from the bulk energy per unit volume integrated over the
finite-size system, and the oscillating contributions come from quantum
interference effects explicitly caused by the confinement.  By supposing that
the smooth part is known while the unknown oscillatory contribution is a
correction, Strutinsky proposed in the 60's a physically motivated approach
to, and an efficient way of calculating, the shell
corrections~\cite{str_str,str_rmp}.

Strutinsky's   shell   correction  method   is   essentially  a   {\it
semiclassical} approximation. It rests on  the fact that the number of
particles  in the  system  considered  is large,  rather  than on  the
interaction between  the particles being  weak. (One must,  of course,
work in a regime where  the smooth starting point is basically valid.)
Since the quantum dots in which we are interested contain on the order
of  a  hundred  electrons, they  are  a  perfect  place to  apply  the
Strutinsky  method.  However,  before   doing  so  for  a  particular,
realistic,  two-dimensional geometry,  we  shall in  this paper  limit
ourselves to a formal discussion  of this method in conjunction with a
one-dimensional  illustrative  example.  In  spite  of the  literature
existing on this subject~\cite{brack}, we  find it useful for two main
purposes.  First,   the  discussion  and   resulting  expressions  are
noticeably  simpler for  quantum  dots than  for  nuclei. This  occurs
because the existence of a smooth confining potential in the dot means
that  gradient corrections  to the  smooth density  are not  needed to
confine    the   system    at   the   zeroth-order  (classical-like)  
approximation.  The effect of  these gradient  terms can  therefore be
included  in the first-order  ``shell'' corrections,  simplifying both
the zeroth-order calculations (no  gradient terms) and the first-order
terms (no corrections to the Weyl part).

Our main purpose, however, is to take advantage of the fact that we use
density functional theory rather than Hartree-Fock as a starting point, the
former being presumably better suited to deal with the long-range Coulomb
interaction present in quantum dots than the latter.  This allows us to
discuss in detail the second-order ``residual interaction'' terms of the
Strutinsky method. By residual interaction we mean the weak interaction
between Landau quasi-particles that comes from dressing the bare electron
added to the quantum dot.  In particular we will show, and illustrate, how
these terms are related to the screened Coulomb interaction.

The remainder of the paper is organized as follows. The Thomas-Fermi and
density functional theories are summarized in the next section, establishing
our notation.  Section~\ref{strutth} contains the Strutinsky method applied
to density functional theory.  This is the core of the paper; in particular,
the relation of the second order terms to the screened Coulomb potential is
derived. Section~\ref{peakspc} recalls how the residual interaction terms
contribute to conductance peak spacing distribution.  Section~\ref{numcalc}
compares the whole approximation scheme to numerical calculations of a
simplified model:  interacting electrons in a one-dimensional quartic
oscillator. Finally, we comment on the relationship between the
Strutinsky development and the constant interaction model, and
possible applications of the  method.

\section{Density Functional Theory}
\label{tflda}

The Hohenberg-Kohn theorem~\cite{ks_hohenberg} states that for a
system of interacting electrons in an external potential,
$\Vext(\br)$, there exists a functional, $\FKS[n]$, of the density of
electrons, $n(\br)$, such that: i)~the density, $n_g(\br)$,
corresponding to the ground state of $N$ particles is an extremum of
$\FKS[n]$ under the constraint that the total number of particles,
\begin{equation} \label{eq:contrainte}
  N[n_g] \equiv \int d\br \, n_g(\br) \; , 
\end{equation}
is fixed, and ii)~$\FKS[n_g]$ is the total energy of the system.  The
explicit form of the Hohenberg-Kohn-Sham functional is not
known~\cite{ks_hohenberg,ks_kohn}. In practice, one must be satisfied
with approximations. We describe here first a generalized Thomas-Fermi
approach and, second, the case when an explicit form of the density
functional is assumed.

\subsection{Generalized Thomas-Fermi Approximation ([Semi]Classical Level)}

It is convenient to view the density functional as the sum of three
parts: a classical charge contribution, the kinetic energy, and the
unknown exchange-correlation functional which accounts for the
balance~\cite{ks_kohn}. The first part is simple: the energy of a
system of classical charges confined by an external potential,
$\Vext$, is \begin{equation} {\cal E}[n] = \Eext[n]+\Ecoul [n]
\end{equation} where \begin{eqnarray} \Eext[n] &=& \int n({\bf r})
\Vext ({\bf r}) d{\bf r}\nonumber\\ \Ecoul [n] &=& {e^2 \over 2}
\int\int {n({\bf r})n({\bf r}') \over |{\bf r}-{\bf r}'|} d{\bf r}
d{\bf r}' \; .  \end{eqnarray}

For the kinetic energy, in the Thomas-Fermi approach the Pauli exclusion
principle is introduced semiclassically by employing the idea that one
quantum state occupies a volume $(2\pi \hbar)^d$ in phase space. This
implies that if many electrons want to be at the same place, they can do so
only by increasing their kinetic energy. This gives the Thomas-Fermi
approximation to the kinetic energy part of the density functional,
$\TTF[n]$, expressed as
\begin{eqnarray}
    && \nu(\epsilon) = {1 \over (2\pi \hbar)^d} 
	\int \Theta \left(\epsilon-{{\bf p}^2 \over 2m} \right) d{\bf p}
	\nonumber \\
    && \tTF(n) = \int_{0}^{n} \epsilon(\nu) d\nu \nonumber\\
    && \TTF[n] = \int \tTF\bigl( n({\bf r})\bigr) d{\bf r} \label{eq:TTF}
\end{eqnarray}
where $d$ is the dimensionality of the system, $\Theta$ is the Heaviside
step function, $\tTF$ is the kinetic energy density, and $\nu(\epsilon)$ is
the number of states per unit volume with energy less than $\epsilon$. A
factor 2 in $\nu(\epsilon)$ is required if the electron spin degeneracy is
taken into account. 

Finally, the effect of exchange and correlation is included through a term
$\Exc [n]$. In practice, an explicit form for this functional must be
taken.  For example, if the electron density is a sufficiently slowly
varying function of position, one can approximate $\Exc [n]$ by taking the
exact results for the uniform electron gas at the local density, the
well-known local density approximation (LDA).

Within this approximation, then, the density functional is
\begin{equation}
    \FTF [n]=\TTF[n]+\Etot[n]
\end{equation}
where 
\begin{equation}
    \Etot[n] \equiv \Eext [n]+ \Ecoul[n]+\Exc [n] \; .
\end{equation}
The ground state energy and its electron distribution are obtained
by minimizing $\FTF$ under the constraint
(\ref{eq:contrainte}), yielding the self-consistency equation 
\begin{equation} \label{eq:TF}
    {\delta \TTF \over \delta n}[\nTF ]({\bf r}) +\Veff [\nTF ]({\bf
    r}) = \mu_{\rm GTF}
\end{equation}
with the effective potential
\begin{equation} \label{eq:Veff}
	\Veff [n]({\bf r}) \equiv
	{\delta \Etot \over \delta n}[n]({\bf r}) \; .
\end{equation}
Notice that to make use of Eq.~(\ref{eq:TF}), one must have an
explicit form for $\Exc$ in order to take the functional derivative in
Eq.~(\ref{eq:Veff}).

We call this approach ``generalized-Thomas-Fermi'' (GTF) because it uses the
Thomas-Fermi approximation for the kinetic energy but retains macroscopic
aspects of exchange and correlation. In particular, the short-range effects 
of exchange-correlation in a uniform system can be included.

\subsection{Kohn-Sham Equations (Quantum Mechanical Level)}

In standard implementations of density functional theory (DFT), the
kinetic energy is treated quantum mechanically rather than
(semi)classically as in GTF.  To accomplish this, one introduces $N$
orthogonal functions $\{\phi_1({\bf r}),...,\phi_N({\bf r})\}$ such
that the electron density is defined as
\begin{equation} \label{denslda}
    n({\bf r}) \equiv \sum_{i=1}^N |\phi_i({\bf r})|^2
\end{equation}
and the kinetic energy is
\begin{eqnarray}
    \TLDA[n] &=& \sum_{i=1}^N \langle \phi_i |\hat T| \phi_i \rangle 
	\nonumber\\
            &=& {\hbar^2 \over 2m} \int 
	\sum_{i=1}^N |\nabla\phi_i({\bf r})|^2 d{\bf r} \; .
\end{eqnarray}
Thus, the density functional becomes
\begin{equation}
    \FLDA[n]=\TLDA [n]+\Etot[n] \; .
\end{equation}

To find the ground state energy, one minimizes $\FLDA$ with respect to
the functions $\phi_i({\bf r})$ under the constraints
\begin{equation}
    \int |\phi_i ({\bf r})|^2 d{\bf r} = 1 \; , \qquad i=1,...,N \; .
\end{equation}
The result is a Schr\"{o}dinger equation 
\begin{equation}
    \label{selda}
    \left(-{\hbar^2 \over 2m} \nabla^2   + 
	\Veff [n]({\bf r}) \right) \phi_i({\bf r}) = 
	\epsilon _i \phi_i({\bf r}) \; , \qquad i=1,...,N
\end{equation}
where $\epsilon _1,...,\epsilon _N$ are Lagrange multipliers and the
effective potential is again defined by Eq.~(\ref{eq:Veff}); these
are the Kohn-Sham equations~\cite{ks_kohn}. The
Eqs.~(\ref{eq:Veff}), (\ref{denslda}), and (\ref{selda}) are the
set of self-consistent equations for finding the electron density,
$\nLDA({\bf r})$, and then the ground state energy $\ELDA = \FLDA[\nLDA]$.

As in the discussion of the GTF above, in order to actually solve the
Kohn-Sham equations, an explicit form for the exchange-correlation
functional is required. The simplest case is when $\Exc$ is an integral
over space of a function ({\it not} functional) of the local density;
this is the well-known local-density approximation (LDA). But other more
complicated explicit forms are possible, for example the generalized
gradient corrections to LDA~\cite{Fulde}.

\section{Approximate Ground State Energy: \\
The Strutinsky Energy-Correction Method}
\label{strutth}

\subsection{Expansion of the DFT Ground State Energy}

In this section we develop an approximation to $\ELDA[\nLDA]$ starting
from the solution of the generalized-Thomas-Fermi equation,
$\nTF$. The main motivation is to develop a physical interpretation of
the difference between these two approaches in finding the ground
state energy. In addition, the approximation is of interest
numerically for large problems since it involves a self-consistent
solution of only the GTF equation rather than the more involved
Kohn-Sham equations. We use the method introduced by
V.~M.~Strutinsky~\cite{str_str,str_rmp} originally in the context of a
Hartree-Fock rather than density-functional approach. His method
describes the interacting system self consistently, first with the
quantum interference effects turned off, and then by introducing them
perturbatively. As discussed in the introduction, the idea is to add
the ``oscillatory'' effects caused by interference in the confined
system to a ``smooth'' essentially macroscopic description---these
effects are essentially the Friedel oscillations~\cite{Ziman} familiar
in the context of impurities or surfaces.

To study the role of quantum interference effects in the DFT ground state
energy, we will first show that the generalized-Thomas-Fermi result is a
[semi]classical approximation to the DFT energy. The GTF approximation does,
of course, contain some quantum mechanics---notably the Pauli exclusion
principle which gives rise to the Fermi surface---and so is not truly
classical. But only the simplest local quantum effects are present in GTF
rather than the effects of interfering paths that one expects in a true
semiclassical theory, hence our characterization of GTF as
``[semi]classical''.

To see this clearly, we introduce a convenient notation adapted from
the semiclassical treatment of single-particle problems: it is
customary there to express the density of states as a sum of a smooth
term slowly varying in energy, called the Weyl part, and a term which
varies rapidly in energy (on the scale of the mean level separation),
called the oscillatory part \cite{gutzwiller}.  For a system governed
by the Hamiltonian $H[V] \equiv {\bf p}^2 /2m + V({\bf r})$, where the
potential is as yet unspecified, one can define the probability
density of $N$ independent particles
\begin{equation}
    n[V]({\bf r}) = \sum_{i=1}^N |\phi _i ({\bf r})|^2
\end{equation}
in terms of the eigenstates $\{ \phi_i \}$ of $H$.  We also define the Weyl
part of $n[V]$ by
\begin{equation} \label{eq:n_W}
    n^W[V]({\bf r}) \equiv {1 \over (2 \pi \hbar)^d} 
    \int \Theta \left[\mu ^W -{\bf p}^2/2m - V({\bf r})\right] d{\bf p}
\end{equation}
where $\mu^W$ must be chosen so that $N=\int n^W ({\bf r}) d{\bf r}$.
Note that $n^W[V]({\bf r})$ is smooth in that it neglects quantum
fluctuations in much the same way that the GTF approximation does.
With this notation, one can derive the useful relation
\begin{equation}
    \label{tfeqproto}
    {\delta \TTF \over \delta n}\left[n^W[V]\right]({\bf r})+V({\bf r}) =
    \mu^W \; .
\end{equation}
Indeed, using 
\begin{equation} \label{eq:dttf_dn}
   \frac{\delta \TTF}{\delta n}[n](\br) = \e\bigl(n(\br)\bigr) \; ,
\end{equation}
Eq.~(\ref{tfeqproto}) reads $\e(n^W[V](\br)) = \mu^W - V(\br)$.
Applying the function $\nu$ introduced in Eq.~(\ref{eq:TTF}) to both sides
of the equality gives the definition of $n^W[V](\br)$,
Eq.~(\ref{eq:n_W}).

Recalling that $\Veff[n]$ is defined as the variational derivative of
$\Etot$ (to be completely clear, it is not the inverse of $n[\Veff]$),
we see that the self-consistency equation (\ref{denslda}) which defines
$\nLDA$ is
\begin{equation} \label{eq:nlda_def_comp}
	\nLDA(\br) = n\bigl[\Veff[\nLDA]\bigr](\br) \; .
\end{equation}
Similarly, Eq.~(\ref{eq:TF}) which defines $\nTF$
can, in applying the above prescription, be put in the form
\begin{equation} \label{eq:ntf_def_comp}
	\nTF(\br) = n^W\bigl[\Veff[\nTF]\bigr](\br) \; .
\end{equation}
These equations do not signify that $\nTF$ is the Weyl part of $\nLDA$;
however, they do indicate that if one neglects the quantum interference
terms (i.e. the difference between the exact particle density and its Weyl
part), then the definitions of $\nTF$ and $\nLDA$ become equivalent. It is
in this sense that $\nTF$ is the [semi]classical approximation of $\nLDA$.

Supposing $\nTF(\br)$ and $\ETF = \FTF[\nTF]$ known, we now
seek to evaluate the corrections to the Thomas-Fermi energy, 
\begin{equation}
   \Delta E \equiv \ELDA - \ETF \; ,
\end{equation}
up to second order in 
\begin{equation}
   \delta n \equiv \nLDA-\nTF \;.
\end{equation}
For this purpose, we first introduce the quantities
\begin{eqnarray}
    \tilde n({\bf r}) &\equiv& n\bigl[\Veff [\nTF ]\bigr]({\bf r})\nonumber \\
    \tilde n^W({\bf r})& \equiv& n^W\bigl[\Veff [\nTF ]\bigr]({\bf r}) 
                                       = \nTF ({\bf r}) \nonumber \\
    \nosc ({\bf r}) &=& \tilde n({\bf r}) - \tilde n^W({\bf
    r}) \; .
\end{eqnarray}
Note that once $\Veff[\nTF]$ is known, all of these can be computed 
through the diagonalization of the known single particle GTF Hamiltonian.
As is well-known, the sum of the eigenvalues of the Kohn-Sham equations,
$\Eop[V] \equiv \sum_1^N \epsilon_i$, does not give the total energy of
the $N$ particles because of double counting of the interaction energy, 
but rather \cite{FHfootnote}
\begin{equation} \label{eq:ELDA}
  \ELDA = \Eop\bigl[\Veff[\nLDA]\bigr] - \int d\br\, \Veff[\nLDA](\br) \, \nLDA(\br)
	+ \Etot[\nLDA] \; .
\end{equation}

To proceed further, we use the relation proved in the appendix
\begin{equation}  \label{append1}
    \Eop[V+\delta V]-\Eop[V]
	\simeq {1 \over 2}\int \delta V({\bf r}) 
	\bigl( n({\bf r})+n'({\bf r}) \bigr) \ d{\bf r}
\end{equation}
where $n(\br) \equiv n[V](\br)$ and $n'(\br) \equiv n[V+\delta V](\br)$
and which is correct through second order in the changes.
Upon inserting $V = \Veff [\nTF ]$ and $ V+\delta V = \Veff [\nLDA] $, 
and thus $n = \tilde n$ and $n' = \nLDA$, the first term on the
right-hand-side of Eq.~(\ref{eq:ELDA}) becomes
\begin{equation} 
    \Eop\bigl[\Veff [\nLDA]\bigr] = \Eop\bigl[\Veff [\nTF ]\bigr]
    + {1 \over 2} \int d{\bf r} \, \delta \Veff ({\bf r})  
	\bigl( \nLDA ({\bf r})+\tilde n ({\bf r}) \bigr) 
\end{equation}
where $\delta \Veff \equiv \Veff [\nLDA] - \Veff[\nTF]$.
Similarly, the second term in $\ELDA[\nLDA]$ is
\begin{eqnarray}
    \int \Veff [\nLDA]({\bf r})\nLDA({\bf r})d{\bf r} &=& 
	\int \Veff [\nTF ]({\bf r}) \, \nTF ({\bf r}) \, d{\bf r} \nonumber\\
    &+& \int \Veff [\nTF ]({\bf r}) \, \delta n({\bf r})   \, d{\bf r} \\
    &+& \int \delta \Veff({\bf r}) \, \nLDA({\bf r}) \, d{\bf r} \; . \nonumber
\end{eqnarray}
Finally, the third term is
\begin{equation}
    \Etot[\nLDA] = \Etot[\nTF ] + 
    \int \bigl( \Veff [\nTF ]({\bf r}) + \delta \Veff ({\bf r})/2 \bigr) 
    \delta n ({\bf r}) d{\bf r}
\end{equation}
with corrections which are third order in $\delta n$.

Combining all the terms together, we obtain
\begin{eqnarray} \label{eq:Edftapprox}
    \ELDA & \simeq & \Eop\bigl[\Veff [\nTF ]\bigr] - 
	\int \nTF ({\bf r}) \, \Veff [\nTF ]({\bf r}) \, d{\bf r} \, + \,  
	\Etot[\nTF ] \nonumber\\
    &+& \case{1}{2} \int \delta \Veff ({\bf r})\,  
                                \tilde n^{\rm osc}({\bf r}) \, d{\bf r} 
   \;.
\end{eqnarray}
In order to express directly the difference between the DFT and GTF
ground states, it is convenient to use 
\begin{equation}
    \ETF [\nTF ]=\TTF[\nTF ]+\Etot[\nTF ] 
\end{equation}
for $\Etot$. In order to simplify the last term in Eq. (\ref{eq:Edftapprox}), 
note that $\nosc$ is of order $\delta n$, and that therefore one
only needs the first order variation of the effective potential, 
$\delta \Veff ({\bf r}) = 
\int (\delta \Veff / \delta n)[\nTF](\br,\br ') \delta n(\br ') d\br '$,
to obtain $\ELDA$ correct through second order. Thus the final
expression for the Strutinsky energy correction is
\begin{equation}
    \ELDA \simeq \ETF +\Delta E^{(1)}+\Delta E^{(2)}
\end{equation}
where the first and second order correction terms are
\begin{eqnarray}
    \Delta E^{(1)} &=& \Eop\bigl[\Veff [\nTF ]\bigr] - 
	\int \nTF ({\bf r}) \, \Veff [\nTF ]({\bf r}) \, d{\bf r} -
	\TTF[\nTF ]
	\label{eq:deltaE1} \\
    \Delta E^{(2)} &=& {1 \over 2} \int \int \tilde 
	n ^{\rm osc}({\bf r}) \, {\delta \Veff  \over \delta n}[\nTF ]
	({\bf r},{\bf r}') \, \delta n({\bf r}') \, d{\bf r} d{\bf r}'  \; .
	\label{eq:deltaE2}
\end{eqnarray}
In this approach, the DFT ground state energy is, then, the sum of a
classical contribution---the generalized-Thomas-Fermi result $\ETF $---and
two quantum contributions---$\Delta E^{(1)}$ and $\Delta E^{(2)}$. 
We now discuss and interpret these two correction terms.

\subsection{Interpretation of the First-Order Corrections}

The first-order correction is simply the oscillatory part of the single
particle energy for a system of $N$ electrons evolving in the potential
$\Veff [\nTF ]$. Indeed, the Weyl part of $\Eop[V]$ is
\begin{equation}
    \Eop^W [V] \equiv {1 \over (2 \pi \hbar)^d}
	\int \Bigl( {{\bf p}^2 \over 2m} + 
	V({\bf r}) \Bigr) \: \Theta \Bigl( \mu ^W - {{\bf p}^2 \over 2m} 
	- V({\bf r}) \Bigr) d{\bf p}d{\bf r}
\end{equation}
where $\mu^W$ is fixed by $N=\int n^W[V]({\bf r})d{\bf r}$. Separately 
integrating the kinetic and potential energy terms for $V=\Veff [\nTF ]$, 
one obtains
\begin{equation}
    \Eop^W \bigl[\Veff [\nTF ]\bigr]= \int d{\bf r}
	\int _0^{\mu^W-\Veff [\nTF ]({\bf r})} 
     \epsilon \: {d \nu \over d \epsilon} d \epsilon 
	+ \int n^W\bigl[\Veff [\nTF ]\bigr]({\bf r}) \: \Veff [\nTF ]({\bf r}) \: d{\bf r}
     \;.
\end{equation}
In the first term one recognizes the Thomas-Fermi kinetic energy, 
$\TTF[\nTF ]$, while in the second term $n^W\bigl[\Veff [\nTF ]\bigr] = \nTF $.
Thus the first-order Strutinsky correction is
\begin{equation}
    \Delta E^{(1)} = \Eop \bigl[\Veff [\nTF ]\bigr]-\Eop^W \bigl[\Veff [\nTF ]\bigr] 
	\equiv \Eop^{\rm osc} \bigl[\Veff [\nTF ]\bigr] \; .
    \label{eq:deltaE1a}
\end{equation}
The leading quantum corrections to GTF are found, then, by quantizing
the single-particle levels in the GTF self-consistent potential:
this is a very natural result which, in fact, was used extensively
in atomic and nuclear physics~\cite{schwinger,mayer,latter,bohigas76} 
before it was first justified by Strutinsky~\cite{str_str,str_rmp}.

\subsection{Interpretation of the Second-Order Corrections}

The second-order correction, Eq.~(\ref{eq:deltaE2}), requires further work:
this form is not useful because it expresses $\Delta E^{(2)}$ as a function 
of the unknown $\delta n$. A second equation is necessary for us to 
determine $\delta n$.  Note that this is not the case for $\Delta E^{(1)}$ 
which is written completely in terms of $\nTF$.

The required second equation is obtained by relating $\delta n$ to 
the oscillatory part of
$\tilde n \equiv n\bigl[\Veff[\nTF]\bigr]$ which, of course, is known since it
depends only on $\nTF$. We start with the two equations
\begin{eqnarray}
	\frac{\delta \TTF}{\delta n}[\nTF] + \Veff[\nTF] & = &
	\mu_{\rm GTF} \label{eq:start1} \\
	\frac{\delta \TTF}{\delta n}[\nLDA^W] + \Veff[\nLDA] & = &
	\mu^W_{\rm DFT} \; . \label{eq:start2}
\end{eqnarray} 
The first equation here is the definition of $\nTF$, and the second one 
follows directly from the general relation (\ref{tfeqproto}). 
Now expand $\nLDA$ about $\nTF$ in the second equation and subtract the
first one from it. In the term involving $\Veff$, $\delta n$ appears.
However, in the kinetic energy term, the density difference is
$\nLDA^W - \nTF = (\nLDA -\nTF) - (\nLDA -\nLDA^W) = \delta n - \nLDAosc $.
To close the equation we must relate $\nLDAosc$ to $\nTF$. This is possible
because in an equation for $\delta n$, which is by definition first
order in corrections, only the first order part of the other quantities
need be kept. Thus, we can approximate $\nLDAosc$ by similarly expanding
$\nLDA$ about $\nTF$, yielding
\begin{equation} \label{eq:sm_nosc}
   \nLDAosc = (\nLDA -\nLDA^W) \simeq (\tilde n - \nTF) = \nosc \;.
\end{equation}
The combination of these results gives the closure equation
\begin{equation} \label{eq:screening}
	\int d\br \frac{\delta \Veff}{\delta n}(\br,\br') \, \delta n(\br')
	+ \int d\br \frac{\delta^2 \TTF}{\delta n^2}(\br,\br')
	\, \bigl(\delta n(\br') - \nosc(\br')\bigr) = \Delta \mu
\end{equation}
where $\Delta \mu \equiv \mu^W_{\rm DFT} - \mu_{\rm GTF}$ is fixed by
the condition $\int \delta n(\br) d\br = 0$.
This is an integral equation for $\delta n$ in terms of GTF quantities.
If a numerical calculation of $\Delta E^{(2)}$ is needed, the
computational cost is relatively modest, largely the inversion of an
operator. 

One obtains a very natural interpretation of the second-order correction
(\ref{eq:deltaE2}) by using this closure equation.
Consider the generalized-Thomas-Fermi problem, Eq.~(\ref{eq:start1}), and
suppose the external potential is slightly modified by the 
quantity $\delta \Vext(\br)$. One thus obtains a new solution of the GTF 
equation $\nTF' = \nTF + \delta \nTF$ which would verify 
\begin{equation}
	\frac{\delta \TTF}{\delta n}[\nTF'] + \Veff[\nTF'] + \delta \Vext = 
	\mu'_{GTF} \; . 
\end{equation}
Subtracting Eq.~(\ref{eq:start1}) as before yields
\begin{equation}
	\int d\br \frac{\delta^2 \TTF}{\delta n^2}(\br,\br')\,\delta \nTF(\br')
	+ \int d\br \frac{\delta \Veff}{\delta n}(\br,\br') \,\delta \nTF(\br')
	+ \delta \Vext = \Delta \mu \; .
\end{equation}
If we now choose the variation of the potential to be
\begin{equation}
	\delta \Vext(\br)   =  \int d\br' \,({\delta^2 \Etot}/{\delta
	n^2})[\nTF](\br,\br')  \,\nosc(\br') \; ,
\end{equation}
$\delta \nTF + \nosc$ satisfies the same Eq.~(\ref{eq:screening}) as 
$\delta n$. This means that, at this level of approximation, $\delta n$ is 
the sum of $\nosc$ and the displacement of charges $\delta \nTF$ screening
$\nosc$ in the GTF approximation. Indeed, the definition of a screened 
interaction $\Vsc$ implies
\begin{equation} \label{eq:Vsc}
	\int d\br' \frac{\delta \Veff}{\delta n}(\br,\br') \,\delta n(\br')
	= \int d\br' \Vsc(\br,\br') \,\nosc(\br') \; ,
\end{equation}
and therefore that the second-order correction, Eq.~(\ref{eq:deltaE2}), 
can be written 
\begin{equation} \label{eq:deltaE2p}
  \Delta E^{(2)} = 
    \frac{1}{2} \int d\br d\br' \,\nosc (\br) 
    \,\Vsc(\br, \br') \,\nosc (\br') \; .
\end{equation}

Thus the second-order correction is simply the energy of interaction between
the additional charge oscillations caused by the quantization, where the
interaction is screened because, after all, the ``other'' electrons treated
in GTF are around. Note that $\Vsc$ is the screened interaction within the
finite sized system, not in the bulk, and so includes boundary
effects~\cite{blanter}; under certain conditions, the bulk screened
potential may be used~\cite{ullmo_unpub}.  More importantly, while the
screened interaction here does include exchange-correlation at the GTF
(macroscopic) level, the result (\ref{eq:deltaE2p}) is a ``direct-like''
contribution while an ``exchange-like'' term is missed. This is related to
the deficiencies of the LDA-like treatment of DFT here and presumably could
be fixed through a local-spin-density functional approach.
Such generalizations would be straight forward using the same arguments
given here.

\section{Contribution of the Residual Interaction to Peak Spacing
Distributions}
\label{peakspc}

As an example of the utility of the Strutinsky method for adding
quantization effects to a macroscopic result, we turn to considering
the spacing between peaks in the conductance through a quantum dot in
the Coulomb blockade regime.  The contribution of residual
interactions have been estimated for chaotic systems within a random
matrix theory framework~\cite{blanter,BroOreHal,BarUllGla,ub_inprep}.
There it was found to be small, but not too far from the scale
necessary to explain the failure of the constant interaction
model. Here our ultimate aim is to evaluate the effect of the residual
interaction in specific model systems which often are not in a regime
where their quantum properties have fully converged to the purely
statistical behavior found in random matrix theory. Systems tend not
to be purely chaotic, and even when chaotic, still exhibit
manifestations of short time dynamics in their eigenproperties. This
can often lead to important deviations from statistical limiting
behaviors. We therefore briefly sketch the relationship between the
residual interaction and the Coulomb blockade peak spacings.

The position of a conductance peak as a function of gate voltage is
proportional to the change in the total energy of the system when an
electron is added \cite{beenakker},
\begin{equation} 
      \mu_N=E(N)-E(N-1) \;, 
\end{equation} 
and the conductance peak spacing is proportional to the discrete
inverse compressibility
\begin{eqnarray} 
	\chi_N &=& \mu_{N+1}-\mu_{N}\\ 
	&=& E(N+1)+E(N-1)-2E(N) \;.  
\end{eqnarray} 
For each of the ground state energies here we will insert the
second-order Strutinsky approximation to the DFT energy.  The first
term, $\ETF$, is the ground state energy in the
generalized-Thomas-Fermi approximation, and is essentially the
charging energy of the dot. The first-order correction contains the
single-particle quantization effects. In some sense these two terms
together constitute the same level of approximation as the much used
constant interaction model. In fact, more physics is included here
since changes in the self-consistent confining potential~\cite{brazil}
are explicitly contained in the Strutinsky approach~\cite{ub_inprep},
whereas due to the ad hoc nature of the constant interaction model,
therein exists no information at all on the self-consistent potential.
The second-order correction term, $\Delta E^{(2)}$, contains, then,
the effects of the residual interaction.

\section{The Quartic Oscillator: A Case Study}
\label{numcalc}

Let us now illustrate the above approach with a particular example.  For the
sake of simplicity, we choose a one-dimensional model system consisting of
$N$ electrons in the confining potential $\Vext(x) = x^4/2$ with the
interactions governed by the one-dimensional Poisson equation ${d^2V_{\rm
int} [n](x) / dx^2} = -4\pi e^2 n(x)$.  This is a simple limit of a
three-dimensional problem: the system is assumed to be invariant in the
tranvserse directions $y$ and $z$ so that the interactions are between planes
of charge, but the medium is extremely inhomogeneous with the transverse mass
taken to infinity so that only one-dimensional quantum mechanics is needed.
Exchange and correlation effects are turned off; thus the interaction
functional is
\begin{equation}
    V_{\rm int}[n](x) = -2\pi e^2 \int_{-\infty}^{\infty}n(x')|x-x'|dx'
   \;.
\end{equation}
Note that use of the 1D Poisson equation causes an interaction which grows
with distance.  Use of the subscript ``int'' in this section, rather than
``coul'' above, is meant to distinguish this case from the three-dimensional
Coulomb interaction.  We emphasize that our interest in this simple model
system is only as an illustration for better understanding of the
Strutinsky method.

We vary the electron charge $e$ to see how well the Strutinsky scheme works
for different strengths of the interaction, $e = 0.5, 1.0$ and $1.5$ in units
where $\hbar=m=1$. The electron spin degeneracy is not considered here.
First,  we perform generalized-Thomas-Fermi and density-functional-theory
calculations directly.  Next, using the GTF results, we apply the Strutinsky
techniques to find approximate DFT results.  Finally, these approximate
results are compared to the actual DFT values.  Because of the neglect of
exchange-correlation here, the GTF approach reduces to true Thomas-Fermi and
the DFT approach is simply the coupled Schr\"{o}dinger-Poisson equations.

\subsection{Thomas-Fermi Numerical Calculations}

For one-dimensional systems
\begin{equation}
    \nu (\epsilon)={1\over2\pi \hbar}\int_{-\infty}^{\infty}
	\Theta (\epsilon -p^2/2m)dp={\sqrt {2m\epsilon}\over \pi
	\hbar}
\end{equation}
and, thus, the kinetic energy term, Eq.~(\ref{eq:TTF}), can be written
explicitly as
\begin{equation}
    \TTF[n] = {\pi ^2 \hbar ^2 \over 6m} \int_{-\infty}^{\infty}n(x)^3
    dx \; .
\end{equation}
The ground state density is obtained by solving the Thomas-Fermi 
equation [cf.\ Eq.~(\ref{eq:TF})]
\begin{equation}
    {\pi ^2 \hbar ^2 \over 2m} \nTF^2(x)
	+\case{1}{2} x^4+V_{\rm int}[\nTF](x)=\mu _{\rm GTF}
\end{equation}
where   we  have used  $(\delta T_{\rm TF} /
\delta  n)[n]=\epsilon(n)=  (\pi^2  \hbar^2 /2m)  n(x)^2$.  By
differentiating twice  and using  the Poisson equation, one  obtains the
second order differential equation
\begin{equation}
    \label{tfde}
    {\pi ^2 \hbar ^2 \over 2m} {d^2 \nTF^2(x) \over dx^2}
    +6x^2-4\pi e^2 \nTF(x)=0 \; .
\end{equation}
This can then be transformed into coupled first-order equations
\begin{eqnarray}
    y_1(x)&=&\nTF^2(x)\nonumber\\
    {dy_1(x)\over dx}&=&y_2(x)\nonumber\\
    {dy_2(x)\over dx}&=&{2m\over \pi^2\hbar^2}\left(4\pi
e^2\sqrt{y_1(x)}-6x^2\right)
\end{eqnarray}
which can be conveniently solved.
Because of the symmetry of  the system, $dn/dx=0$ at the origin and one
need only specify the density at the origin as an initial condition. One 
repeats solving  Eq.~(\ref{tfde}) adjusting $\nTF(x=0)$ on  each iteration  
until the normalization  condition $N=\int  \nTF(x)dx$ is  satisfied.
Once  the electron  density  $\nTF(x)$ is  found,  the ground  state
energy is obtained from
\begin{equation}
   \ETF = \TTF[\nTF]+\Eext[\nTF]+\Eint[\nTF]
\end{equation}
where $\TTF$ is given in Eq~(\ref{tfde}) and
\begin{eqnarray}
    \Eext[\nTF] &=& 2\int_0^{\infty}\nTF(x){1\over 2}x^4dx \nonumber\\
    \Eint[\nTF] &=& 2\cdot {1\over 2}
	\int_0^{\infty}\nTF(x)V_{\rm int}[\nTF](x)dx \;.
    \label{quarticEextEcoul}
\end{eqnarray}

The electron densities $\nTF$ for $N=5, 10$ and $20$ with $e=1.0$ are
plotted in Fig.~1(a) and the effective potential, $\Veff[\nTF]$ given by
Eq.~(\ref{eq:Veff}), in Fig.~1(b).  All three cases show the same basic
structure which can be simply understood as follows.  Without the
interaction ($e=0$), the density would have one maximum at the origin since
the external potential has a minimum at the center.  Once the interaction is
turned on, electrons repel each other and avoid the center, making two
maxima in the density.  Though not pictured, the larger the value of the
electron charge $e$, the lower the central valley in the density, and the
more the density maxima move away from the origin.  As intuitively expected,
the minimum points in the effective potential correspond to the maximum
points of electron density, and increasing $e$ increases rapidly the bimodal
nature of the density.

\subsection{Quantum Numerical Calculations}

The numerical calculation of the DFT energy requires self-consistently
solving
\begin{eqnarray}
    && \left( -{\hbar^2 \over 2m} {d^2 \over dx^2} + 
        \Veff[n](x) \right) \phi_i(x) = \epsilon _i \phi_i(x)\nonumber\\
    && n(x) = \sum_{i=1}^{N} |\phi_i(x)|^2\\
    && \Veff[n](x) = \case{1}{2} x^4 - 2\pi e^2 
       \int_{-\infty}^{\infty}n(x') |x-x'| \ dx' \nonumber
\end{eqnarray}
which are the coupled Schr\"{o}dinger-Poisson equations.
We start the self-consistent iterations with the Thomas-Fermi potential
$\Veff[\nTF]$.  At each iteration, we first diagonalize the Hamiltonian
$\hat H  = {\bf \hat p}^2 / 2m  + \Veff(\br)$ expressed in the  basis of
$\hat  H_0  =  {\bf \hat  p}^2/2  +  x^4/2$.   From the eigenvalues and
eigenvectors of $\hat  H \bigl[\Veff[n]\bigr]$, we can construct the  electron
density  and  the  corresponding  effective potential.  Self-consistency is
evaluated by  comparing the effective potentials $\Veff^{\rm Old}$ and
$\Veff^{\rm  New}$ before and after each iteration (or equivalently the
densities $n^{\rm  Old}$ and $n^{\rm New}$).

Because of the well-known instability of the Poisson equation, one cannot
simply use the output from one iteration, $\Veff^{\rm New}$, as the input to
the next~\cite{hf_herman}.  Instead, we feedback only part of the output
\begin{equation}
    \Veff^{\rm Old,~Next~Iteration}=\Veff^{\rm Old}+\alpha
    (\Veff^{\rm New}-\Veff^{\rm Old}), \; 0<\alpha<1
\end{equation}
where  $\alpha$ is initially  set as  $0.5$. If  the self-consistency is not
improved, $\alpha$ is reduced  by half and the  iteration redone so that
improvement is guaranteed for every iteration. We repeat this until the
density reaches self-consistency,
\begin{equation}
    \max_{|x|\leq x_{\rm max}}|n^{\rm New}(x)-n^{\rm OLD}(x)|\leq 10^{-9}.
\end{equation}
We require self-consistency in the density rather than the potential because
the overall magnitude of the density does not change significantly as $N$
increases.

Once the self-consistent density and effective potential are obtained, the
Weyl part of the density, $n_{\rm DFT}^W$, as well as the chemical potential
$\muLDAW$ can be calculated from Eq.~(\ref{eq:n_W}); the oscillating part of
the density follows from $\nLDAosc=\nLDA-\nLDAW$.  Finally, the
self-consistent ground state energy is obtained using
\begin{equation}
    \TLDA[\nLDA] =  \Eop\bigr[\Veff[\nLDA]\bigr]
-2\int_0^{\infty}\nLDA(x)\Veff[\nLDA](x)dx
\end{equation}
and the same expressions for $\Eext$ and $\Eint$ as in the Thomas-Fermi
calculation, Eq.~(\ref{quarticEextEcoul}).  We have used the above relation
for the kinetic energy instead of the definition since the eigenvalues are
more stable than the eigenvectors in the numerical calculations.

The quantum electron densities for $N=5, 10$, and $20$ are superposed in
Fig.~1(a) for electron charge $e=1.0$.  One can see the quantum mechanical
oscillations whose number of peaks corresponds to the electron number $N$.
Note the decreasing oscillation amplitudes with increasing particle number,
as well as the tunneling outside of the potential wall at the classical
turning points.  The effective potentials, superposed in Fig.~1(b), are
indistinguishable from the corresponding Thomas-Fermi potentials.

\subsection{Strutinsky Energy Corrections}

In order to find the approximate ground state energy using the Strutinsky
method, we start with the Thomas-Fermi density and potential, calculated
above, and quantize in this potential by solving the Schr\"{o}dinger
equation $\left(-{\hbar^2\over 2m}{d^2\over  dx^2} + \Veff[\nTF]\right)
\phi_i =  \epsilon_i \phi_i$ for the eigenvalues and eigenvectors.

The first-order energy correction is given by Eq.~(\ref{eq:deltaE1a});
in our example, the expression for the Weyl part reduces to
\begin{eqnarray}
    {\cal E}_{\rm 1p}^W\bigr[\Veff[\nTF]\bigr]
    = 2{\sqrt{2m}\over 3\pi\hbar}
    \int_0^{\infty}\bigl(\mu^W+2\Veff[\nTF](x)\bigr) \,
    \sqrt{\mu^W-\Veff[\nTF](x)} \, dx \;.
\end{eqnarray}

The second-order energy correction, from Eq.~(\ref{eq:deltaE2}), is
\begin{eqnarray}
    \Delta E^{(2)} &=& -\pi e^2\int_{-\infty}^{\infty}\int_{-\infty}^{\infty}
                       \nosc(x)|x-x'|\delta n(x')dxdx'\nonumber\\
    &=&4\pi e^2\int_{0}^{\infty}\nosc(x)
       \left\{ \int_{x}^{\infty}(x-x')\delta n(x')dx'\right\} dx \;.
\end{eqnarray}
The required input $\delta n$ follows from Eq.~(\ref{eq:screening}).  This 
equation can be simplified by noting, first, for the kinetic energy term
\begin{equation}
    {\delta^2\TTF\over \delta n^2}[\nTF](x,x') = {\delta \epsilon
   (\nu)\over \delta \nu}[\nTF] \, \delta(x-x') \;.
\end{equation}
Second, for the term depending on $\Veff$ note that
\begin{equation}
    \int_{-\infty}^{\infty}{\delta \Veff\over \delta n}[\nTF](x,x')\delta
n(x')dx' = -2\pi e^2\int_{-\infty}^{\infty}\delta n(x')|x-x'|dx'=V_{\rm
int}[\delta n](x)
\end{equation}
implies
\begin{equation}
    {d^2V_{\rm int}[\delta n](x)\over dx^2}  =
               -4\pi e^2\delta n(x) \;.
\end{equation}
Thus, by taking the second derivative with respect to $x$ of 
Eq.~(\ref{eq:screening}), one obtains
\begin{equation}
    \label{deltande}
    {\pi^2\hbar^2\over m}{d^2\over dx^2}\bigl\{ \nTF(x) \cdot 
    \bigl( \delta n(x)-\nosc(x) \bigr) \bigr\} - 4\pi e^2\delta n(x)=0
    \;.
\end{equation}
This equation can be converted into the coupled first-order equations
\begin{eqnarray}
    y_1(x)&=&\nTF(x)\cdot
           \left( \delta n(x)-\tilde n^{\rm osc}(x) \right) \nonumber\\
    {dy_1(x)\over dx}&=&y_2(x)\nonumber\\
    {dy_2(x)\over dx}&=&{4me^2\over \pi\hbar^2}\left({y_1(x)\over
\nTF(x)}+\tilde n^{\rm osc}(x)\right)
\end{eqnarray}
which can be conveniently solved.

With the energy correction terms calculated, the Strutinsky scheme allows us
to approximate the quantum ground state energy using essentially classical
Thomas-Fermi quantities.  We plot $\Delta E\equiv \ELDA-\ETF$, $\Delta
E-\Delta E^{(1)}$, and $\Delta E-\Delta E^{(1)}-\Delta E^{(2)}$ as functions
of $N$ to see the series convergence of the Strutinsky scheme in Figs.~2,
3 and 4 for $e=0.5, 1.0$ and $1.5$ respectively. In the first two cases,
the convergence seems to be working well: for $e=0.5$, without correction
terms the error is of order 0.1 and smooth, while the first-order correction
term improves the accuracy to 0.001, and the second-order term to roughly
0.0001.  For $e=1.0$, without correction terms the error is about 0.1, with
the first-order correction the error reduces to 0.01, and with the
second-order to 0.0004.

For $e=1.5$, before comparing the order of magnitude of the different terms,
it is useful to say a few words about the odd-even structure clearly visible
in Fig.~4, and actually already apparent for $e=1.0$ at low $N$ in
Fig.~3. The origin of this behavior is not related to spin (which has not
been included) but can be readily understood by looking at the lower panel
of Fig.~5, which shows the effective Thomas-Fermi potential $\Veff[\nTF]$
for $N=20$, $e=1.5$. Here, one sees that the latter has developed a barrier
at the center of the well, the top of which lies very close to the chemical
potential. For the quantum case, this means that the one particle levels
below the Fermi energy are organized as quasi-doublets.  This naturally
leads to an odd-even effect since for N even (odd), the last occupied
orbital has an energy consistently below (above) that suggested by the Weyl
approximation.

Moreover, because the last occupied orbitals are close to the chemical
potential and so near the top of the barrier, it is clear that semiclassical
approximations will be ``at risk'' here. This is clearly seen for instance
in Figs.~6 and 7, which, for coupling $e=1.5$ and $N=19$ and $20$ particles,
shows a comparison between the exact $\delta n(\br) \equiv \nLDA(\br)
-\nTF(\br)$ and its approximation obtained using Eq.~(\ref{eq:screening}).
The two curves are almost on top of each other everywhere, except in the
middle of the well---that is, near the maxima of the barrier. In addition,
one observes that in that place, the approximation is worse for an odd than
for an even number of particles. This can be explained by the fact that in
the former case the last occupied orbital is symmetric and thus has a
probability maximum at the origin, while in the latter case the last
occupied orbital is antisymmetric and so goes to zero. As a consequence the
errors in the Strutinsky approximation scheme also display an odd-even
structure clearly seen in Fig.~5. If, however, one concentrates on the
even case, where the effect of the central barrier is less important,
without correction terms the error is about 0.2, with the first-order
correction the error is reduced to 0.02, and further reduced to about 0.003
if one includes the second-order corrections.

\subsection{The Peak Spacings}

Turning now to the inverse compressibility $\chi_N$ introduced in
Sect.~\ref{peakspc}, we observe the same trend as for the groundstate
energy. The approximation $\chi_N^{\rm STR(0)}$, calculated strictly
by the Thomas-Fermi approximation, already gives an excellent relative
precision.  Moreover, for $e=0.5$ and $e=1.0$, each term in the
Strutinsky development significantly reduces the error (i.e.~an order
of magnitude or more for $N=18$).  For these cases however the
approximation is already much better than the mean spacing $\Delta$ if
the first correction is included.  We therefore show the data only for
$e=1.5$, for which the corrections are enhanced by the proximity of
the chemical potential to the top of the inner barrier.  In Fig.~8(a),
the $\chi_N$ are shown.  The discrete points represent the full
quantum calculations and are taken as the reference points. It is seen
that $\chi_N^{\rm STR(0)}$ does not capture the odd-even
double-well effect, but otherwise captures the essential peak spacing
behavior.  In Fig.~8(b), the relative errors are shown as a function
of $N$.  More specifically, the difference between the quantum
$\chi_N$ and one, two, or three terms of the Strutinsky
series---$\chi_N^{\rm STR(0)}$, $\chi_N^{\rm STR(1)}$ and $\chi_N^{\rm
STR(2)}$, respectively---is divided by $\Delta$, the mean level separation.  
It is
seen that the majority of the odd-even, double-well effect is captured
by the first correction term.  Reassuringly, even in this case for
which the inner barrier degrades the quality of the semiclassical
development, each inclusion of an extra term in the series reduces
somewhat the relative errors of the Strutinsky method.  Moreover the
improvement due to the addition of the second order correction becomes more
significant  with increasing $N$.  Thus we see that even in this 
more difficult case,  the Strutinsky method gives a excellent
scheme of approximation in the semiclassical limit.

\section{Conclusion}

We have developed an approximate series expansion for studying the
ground state of interacting systems using the idea of the Strutinsky
shell correction method. We tested the validity of the Strutinsky
scheme by numerical calculations of interacting electron systems in a
one-dimensional, externally applied quartic potential. By varying the
electron charge strength, we were able to confirm the stability of the
method. It approximated extremely well the quantum mechanical DFT
quantities using [semi]classical Thomas-Fermi data for three different
charge strengths.  One exceptional circumstance giving less reliable
results was noted with respect to a barrier in $\Veff$ approaching the
chemical potential.  The calculations show a tendency for the series to
converge as electron number increases.  This is consistent with
expectations for systems with large numbers of particles because of
the increasing reliability of the Thomas-Fermi calculations as the
system goes deeper into the semiclassical regime.

The method discussed in this paper could serve two purposes. On one hand, it
gave us an efficient way to proceed with numerical calculations, and it is
conceivable that this approach could be of some help for larger scale,
realistic DFT calculations~\cite{stopa,wingreen,Fulde}.  On the other hand,
it also provides some physical insight by decomposing the total DFT energy
into various contributions, each of them receiving an intuitive
interpretation.

In the context of quantum dots, for instance, one of its simple but
presumably useful applications could be to justify, and make more
precise, the constant interaction model. Indeed, this latter is
usually presented as an ad-hoc model motivated essentially by its
simplicity. Here, up to some reinterpretation of what is the
capacitance of the dot, we see that the constant interaction model can
be understood as the first-order approximation in a Strutinsky
development of a DFT calculation. One obtains, in addition, that the
various parameters of the model (classical energy and potential
governing the motion of the independent particles) are specified, and,
in principle, can be computed explicitly. This makes it possible, for
instance, to study the sensitivity of the dot's energy to the
variation of an external parameter \cite{ub_inprep}.

In the same way, the second-order correction term gives insight into the
``residual'' interactions between electrons. In the context of DFT it gives
some basis to the fact that electrons in quantum dots behave as Landau
quasi-particles interacting weakly through a screened Coulomb interaction. It
moreover explicitly specifies how the screening process is affected by the
confinement of the electrons, which might be relevant in the limit where the
screening length is not much smaller than the size of the dots.

With the failure of the theoretical predictions of the single-particle
constant-interaction model to explain the experimentally observed conductance
peak spacing
distributions~\cite{peakspc_sivan,peakspc_patel,peakspc_simmel1,peakspc_simmel2,peakspc_ensslin},
our original interest was to investigate what physical factors are missing
and to understand better the statistical behaviors of quantum dots.  We
expect that the approach of defining the many-body system, explicitly
including electron-electron interactions within density functional theory, to
bring us closer to a resolution of the failure of the constant interaction
model in this context.  We leave application to more realistic systems
for future work.

\section*{Acknowledgments}

We acknowledge helpful conversations with Oriol Bohigas, Matthias
Brack, Nicholas Cerruti and Nicolas Pavloff.
The LPTMS is an ``Unit\'e de recherche de l'Universit\'e Paris 11
associ\'ee au C.N.R.S.''
This work was supported in part by NSF grant \#PHY-9800106

\appendix

\begin{section}{}

Consider a Hamiltonian $H={\bf p}^2/2m+V({\bf r})$ and its perturbed
Hamiltonian $H'=H+\delta V({\bf r})$. We denote the eigenvalues and
eigenvectors as $\epsilon _i$ and $\phi _i({\bf r})$ [$\epsilon _i'$ and
$\phi _i'({\bf r})$] for $H$ ($H'$).

To second order in $\delta V$, the perturbed eigenvalues are
    \begin{equation}
    \label{epsp}
    \epsilon _i' = \epsilon _i + \epsilon _i^{(1)} + \epsilon _i^{(2)}
    \end{equation}
where 
    \begin{eqnarray}
    \epsilon _i^{(1)} &=& \langle \phi_i|\delta V|\phi_i\rangle\nonumber\\
    \epsilon _i^{(2)} &=& \sum_{j \neq i} {|\langle \phi_i|\delta
V|\phi_j\rangle|^2 \over \epsilon_i-\epsilon_j} \;, 
    \end{eqnarray}
assuming non-degenerate eigenstates.

Similarly, taking $H'$ as the original Hamiltonian and $H=H'-\delta V$ as
the perturted Hamiltonian, one can write
    \begin{equation}
    \label{eps}
    \epsilon _i = \epsilon _i' + \epsilon _i'^{(1)} + \epsilon _i'^{(2)}
    \end{equation}
where 
    \begin{eqnarray}
    \epsilon _i'^{(1)} &=& -\langle \phi_i'|\delta
V|\phi_i'\rangle\nonumber\\
    \epsilon _i'^{(2)} &=& \epsilon _i^{(2)}+o(\delta V^3) \;.
    \end{eqnarray}
Subtracting Eq.(\ref{eps}) from Eq.(\ref{epsp}), one obtains to
second order in $\delta V$
    \begin{eqnarray}
    \epsilon _i'-\epsilon _i &=& {1 \over 2}\left(\epsilon
_i^{(1)}-\epsilon _i'^{(1)}\right)\nonumber\\
    &=& {1 \over 2}(\langle \phi_i|\delta V|\phi_i\rangle+\langle
\phi_i'|\delta V|\phi_i'\rangle) \;.
    \end{eqnarray}
Summing over $i$, one obtains
    \begin{eqnarray}
    {\cal E}_{1p}[V+\delta V]-{\cal E}_{1p}[V] &=& {1 \over
2}\sum_{i=1}^N\left(\langle \phi_i|\delta V|\phi_i\rangle+\langle
\phi_i'|\delta V|\phi_i'\rangle\right)\nonumber\\
    &=& {1 \over 2} \int \delta V ({\bf r})(n({\bf r})+n'({\bf r}))d{\bf
r} \;.
    \end{eqnarray}

\end{section}

\begin{figure}
\begin{center}
\leavevmode
\epsfxsize = 8.6cm
\epsfbox{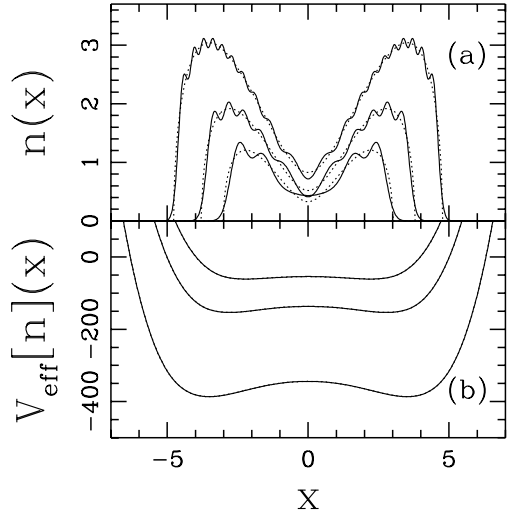}
\end{center}
\caption{
The electron densities (upper panel) and effective potentials (lower panel)
for interacting particles in a quartic potential.  The results
for both the Thomas-Fermi (dashed) and coupled Schr\"odinger-Poisson (solid)
approximations are given.  The electron charge is $e=1.0$, and the electron
number is $N=5, 10$, and $20$ from bottom to top in upper panel, top to
bottom in lower panel.  $\Veff[\nTF]$ and $\Veff[\nLDA]$ coincide so well
that one cannot distinguish the differences here.
}
\label{fig1}
\end{figure}

\begin{figure}
\begin{center}
\leavevmode
\epsfxsize = 8.6cm
\epsfbox{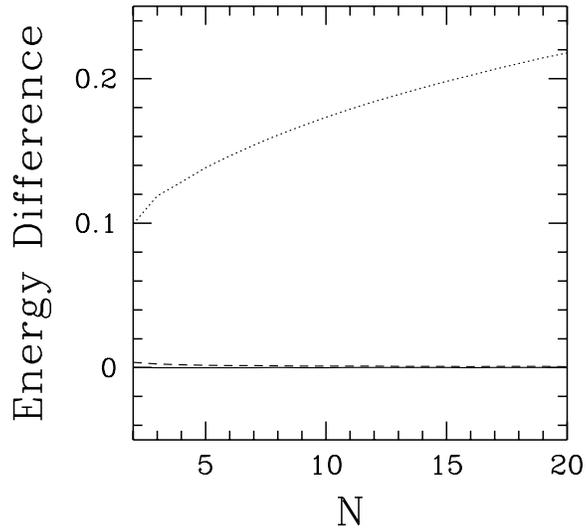}
\end{center}
\caption{
Convergence of approximations to the quantum ground state energy
for electron charge $e=0.5$. The curves are, from top to
bottom, the error in the Strutinsky energy correction approach taken
at zeroth, first, and second order: specifically
$\Delta E \equiv \ELDA-\ETF$  (dotted),
$\Delta E-\Delta E^{(1)}$ (dashed), and
$\Delta E-\Delta E^{(1)} -\Delta E^{(2)}$ (solid).
The convergence in this case is excellent.
}
\label{fig2}
\end{figure}

\begin{figure}
\begin{center}
\leavevmode
\epsfxsize = 8.6cm
\epsfbox{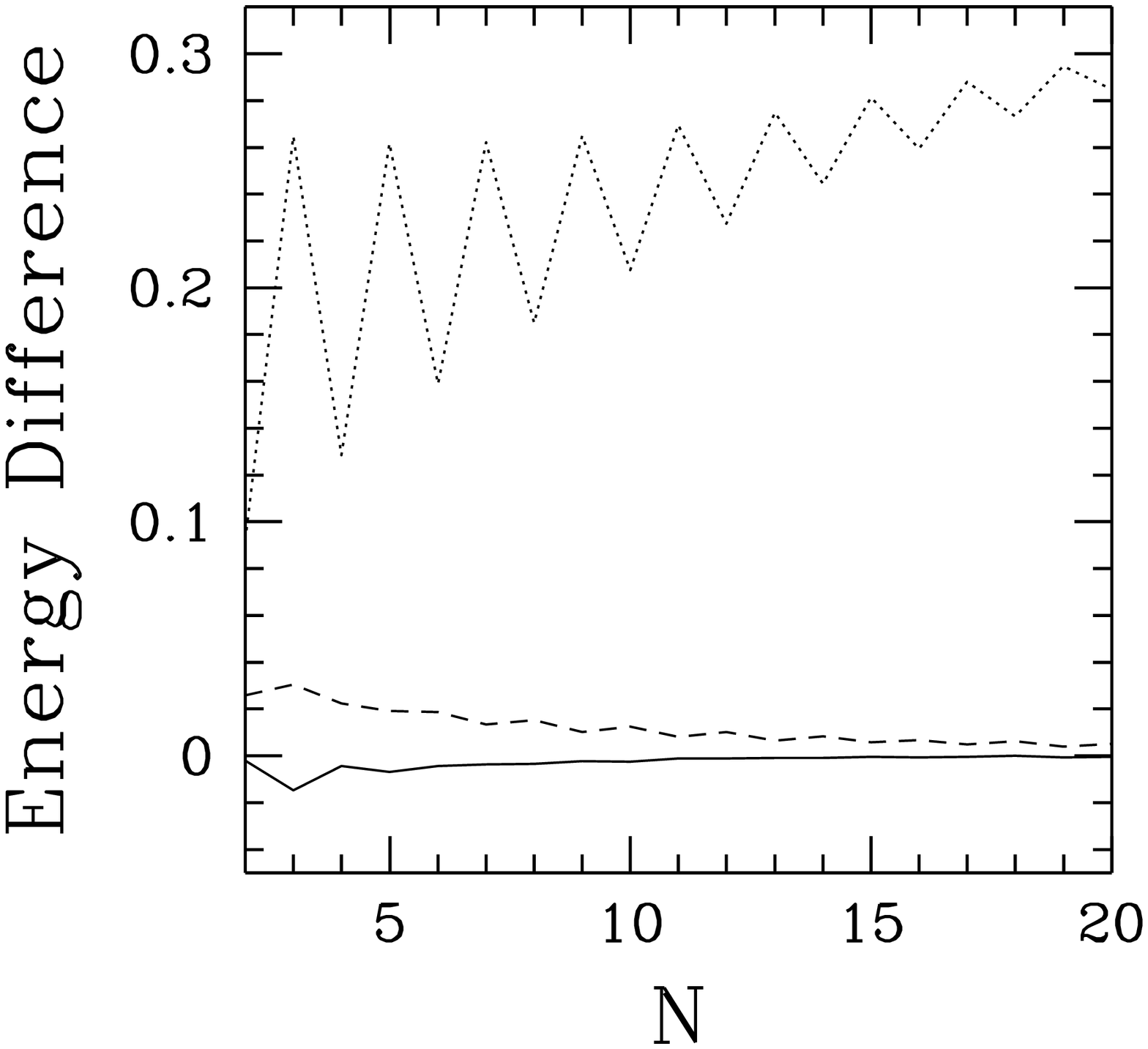}
\end{center}
\caption{
Convergence of approximations to the quantum ground state energy
for electron charge $e=1.0$. The curves are, from top to bottom, 
$\Delta E \equiv \ELDA-\ETF$  (dotted),
$\Delta E-\Delta E^{(1)}$ (dashed), and
$\Delta E-\Delta E^{(1)} -\Delta E^{(2)}$ (solid).
}
\label{fig3}
\end{figure}

\begin{figure}
\begin{center}
\leavevmode
\epsfxsize = 8.6cm
\epsfbox{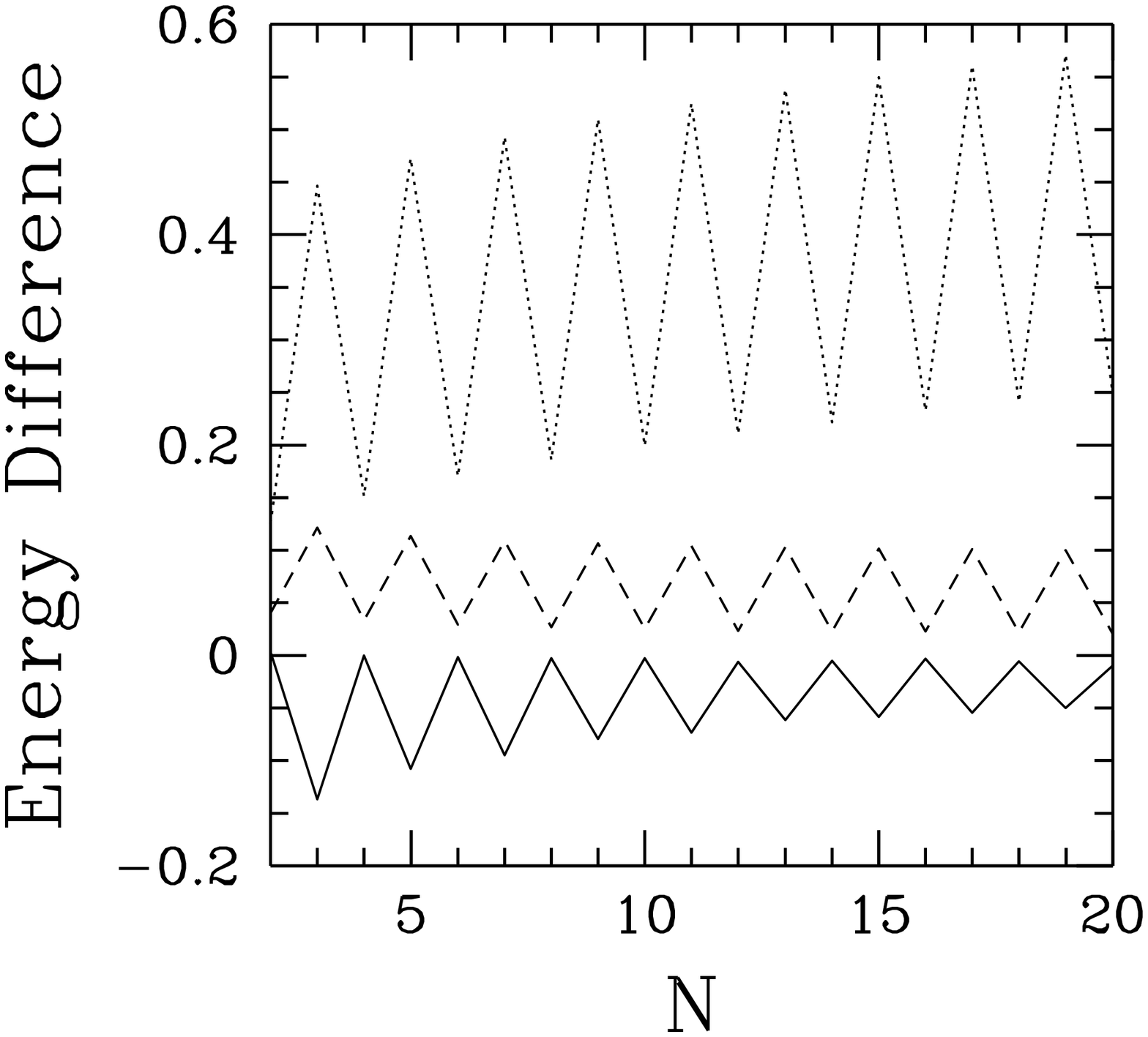}
\end{center}
\caption{
Convergence of approximations to the quantum ground state energy
for electron charge $e=1.5$. The curves are, from top to bottom, 
$\Delta E \equiv \ELDA-\ETF$  (dotted),
$\Delta E-\Delta E^{(1)}$ (dashed), and
$\Delta E-\Delta E^{(1)} -\Delta E^{(2)}$ (solid).
}
\label{fig4}
\end{figure}

\begin{figure}
\begin{center}
\leavevmode
\epsfxsize = 8.6cm
\epsfbox{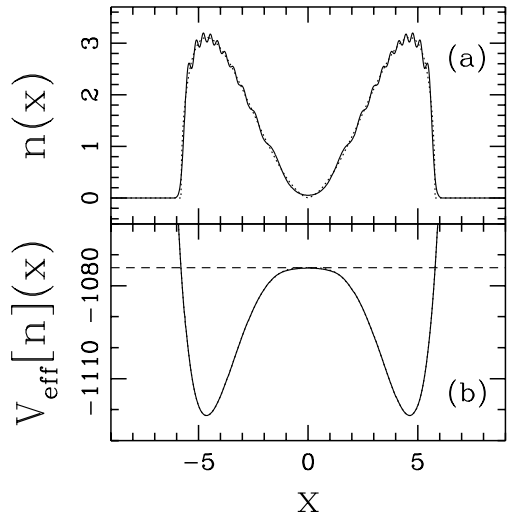}
\end{center}
\caption{
The electron density (upper panel) and effective potential (lower panel) for
$N \!=\! 20$ particles and $e \!=\! 1.5$.  In the upper panel, results for
both the Thomas-Fermi (dashed) and quantum (solid) cases are given.  In the 
lower panel, the dashed horizontal line is the position of the chemical 
potential $\mu_{\rm TF}$.
}
\label{fig5}
\end{figure}

\begin{figure}
\begin{center}
\leavevmode
\epsfxsize = 8.6cm
\epsfbox{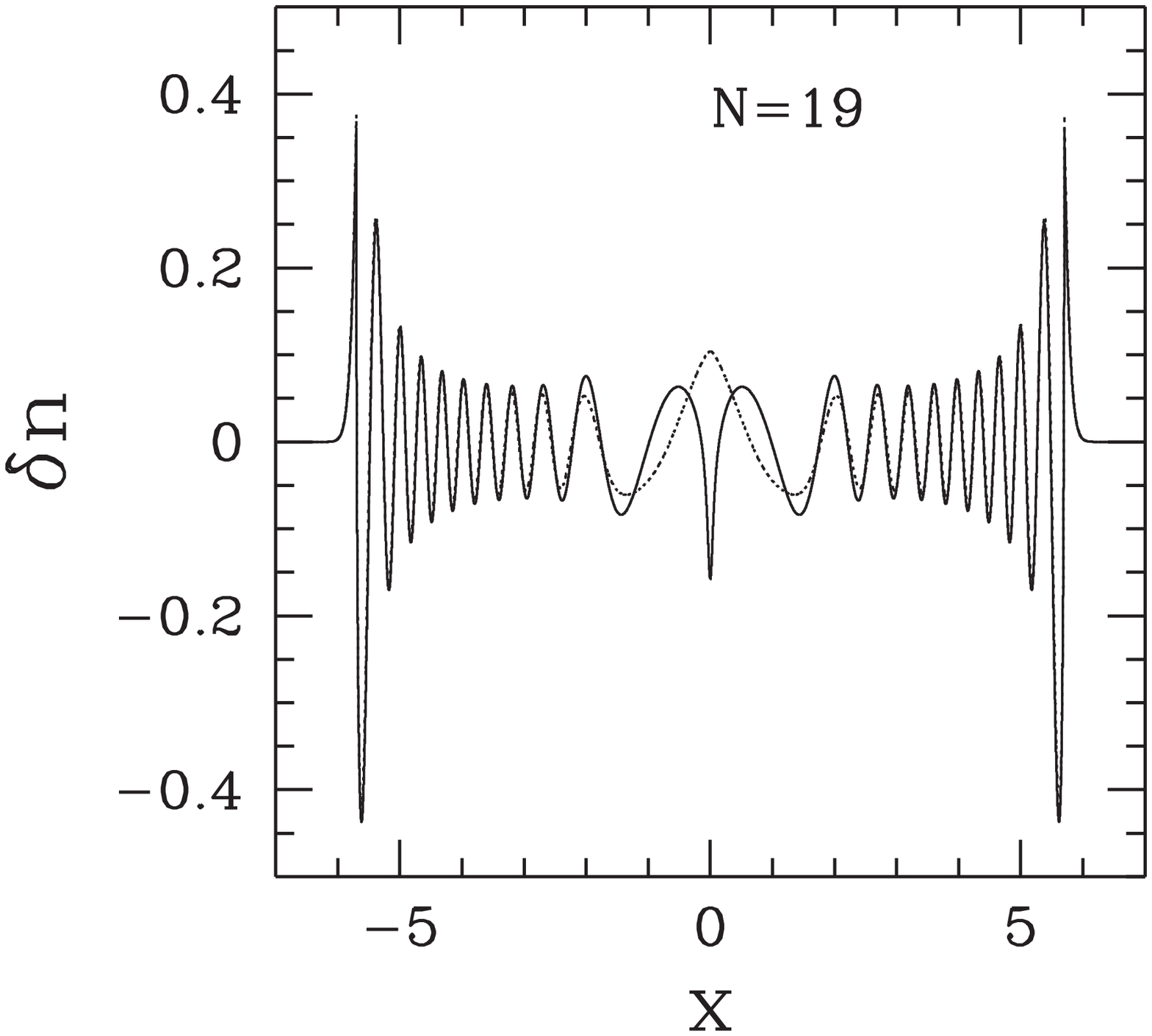}
\end{center}
\caption{
Comparison between the exact $\delta n \equiv \nLDA -
\nTF$ (dashed line) and its approximation derived from
Eq.~(\ref{eq:screening}) (solid line) for $e=1.5$ and $N=19$.
}
\label{fig6}
\end{figure}

\begin{figure}
\begin{center}
\leavevmode
\epsfxsize = 8.6cm
\epsfbox{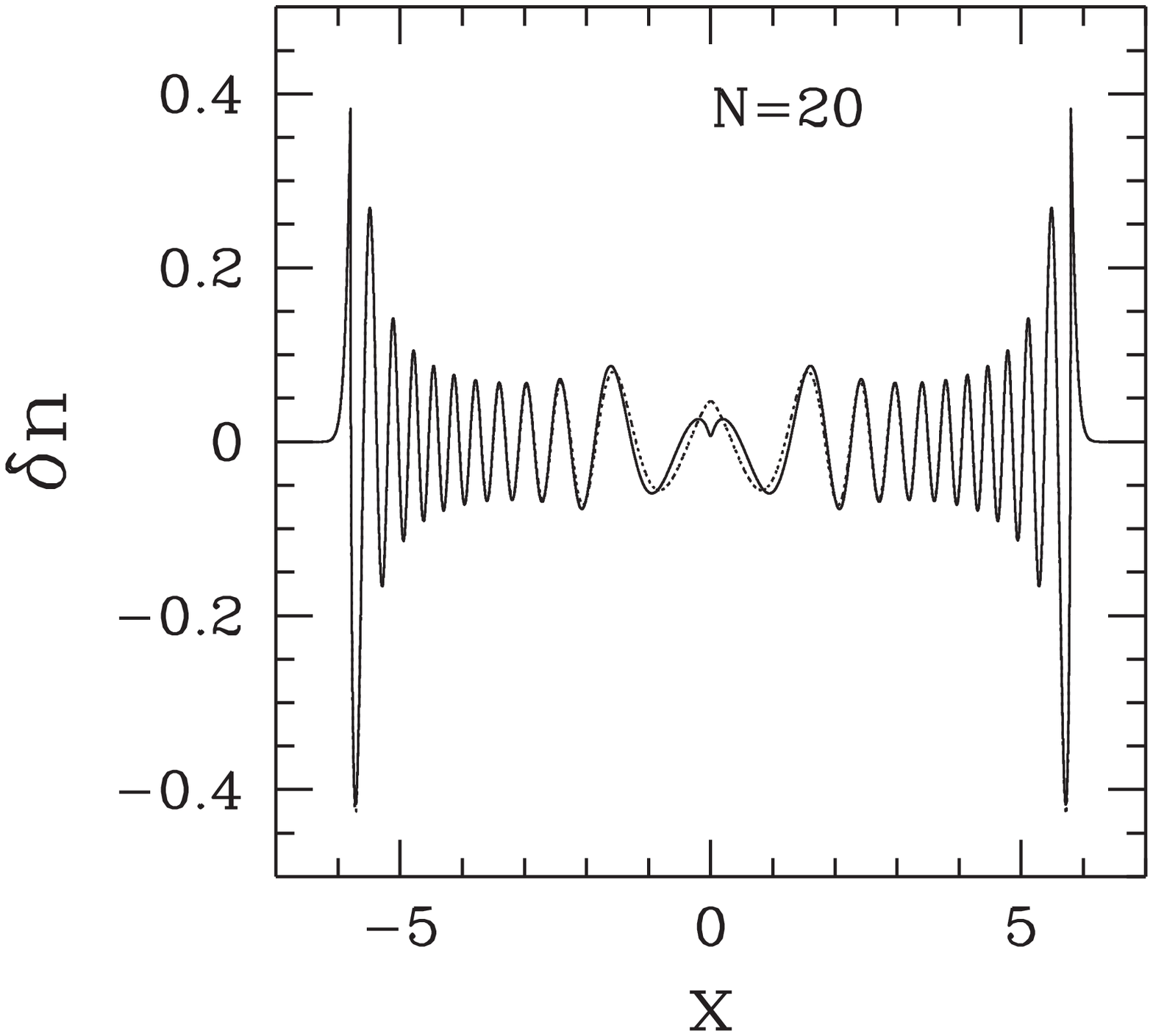}
\end{center}
\caption{
Comparison between the exact $\delta n \equiv \nLDA -
\nTF$ (dashed line) and its approximation derived from
Eq.~(\ref{eq:screening}) (solid line) for $e=1.5$ and $N=20$.
}
\label{fig7}
\end{figure}

\begin{figure}
\begin{center}
\leavevmode
\epsfxsize = 8.6cm
\epsfbox{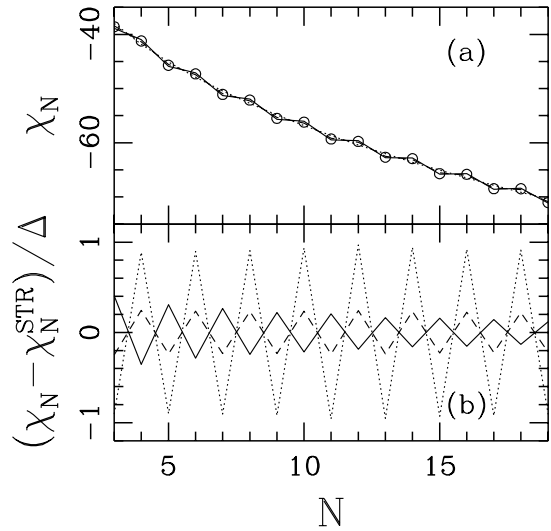}
\end{center}
\caption{
The discrete inverse compressibility as a function of
electron number $N$. The open circles are $\chi_N$
using the quantum ground state energy (exact). Results using three
approximate ground state energies are shown: 
(1)~dotted: Thomas-Fermi, $\chi_N^{\rm STR(0)}$ using $\ETF$,
(2)~dashed: first-order Strutinsky,
$\chi_N^{\rm STR(1)}$ using $\ETF+\Delta E^{(1)}$, and
(3)~solid: second-order Strutinsky,
$\chi_N^{\rm STR(2)}$ using $\ETF+\Delta E^{(1)}+\Delta E^{(2)}$. 
The lower panel shows the relative errors, $(\chi_N-\chi_N^{\rm
STR})/\Delta$, of the same three approximations.
}
\label{fig8}
\end{figure}

\end{document}